\documentclass[12pt]{article}
\usepackage[applemac]{inputenc}\usepackage[english]{layout}
\usepackage[english]{babel}
\usepackage{multicol}
\usepackage{ae,aecompl}
\usepackage{fancyhdr}
\usepackage{amsmath,amssymb,latexsym, cancel, feyn}
\usepackage{amssymb,latexsym}
\usepackage{graphicx}
\usepackage{array}
\usepackage[small,sf]{caption}
\usepackage[headheight=10pt,headsep=15pt,footskip=1.5cm,hmargin=2.0cm,includeheadfoot,top=2cm,bottom=1.5cm]{geometry}
\newsavebox{\mygraphic} 

\title{Geometrical Approximation to the AdS/CFT Correspondence}
\author{M. A. Martin Contreras\thanks{ma.martin41@uniandes.edu.co}, J. M. R. Roldan Giraldo\thanks{jroldan@uniandes.edu.co}\\ \small High Energy Group, Department of Physics, University of los Andes}
\date{}

\begin{document}
\maketitle

\begin{abstract}
In this paper an analysis of the geometrical construction of the AdS/CFT Correspondence is made. A geometrical definition of the configuration manifold and the boundary manifold in terms of the conformal compactification scheme is given. As a conclusion, it was obtained that the usual definition of the correspondence \cite{malda1} is strongly dependent of the unicity of the conformal class of metrics on the boundary. Finally, a summary of some of the geometrical issues of the correspondence is made, along with a possible way to avoid them.
\end{abstract}

\section{Introduction} 
Gravity/Gauge duality is maybe one of the most important developments of the latest times in String Theory. From its very begining, dual models have been applied in many areas different from High Energy Physics or Black Hole Physics. Any branch of Physics that exhibits phase transitions can be modeled using dual models\cite{papa}. 

The central idea of Gravity/Gauge duality is the geometrical connection existing between any Gravity Theory (Superstrings, for example) in $d+1$ dimensions to a QFT living in $d$ dimensions. In fact, it can be said that \emph{we can extract information about QFT from spacetime}, and viceversa. This is just a conjecture, and it still needs a proof. Once the connection between bulk and boundary is stablished, the next step is to write of a proper holographic dictionary, allowing to switch between gravity and QFT.

AdS/CFT Correspondence \cite{malda1} is the most relevant realization of the Gravity/Gauge duality, but is not the only successful one. Some examples of this kind of duality are the Klebanov--Strassler duality \cite{klev1}  or the NS5--branes/LST \cite{ahar1}. In all the three cases mentioned above, the bulk is a non--compact manifold endowed with gravity, such that the dual gauge theory is encoded in its asymptotic behavior. 
\section{AdS/CFT Correspondence in a Nutshell} 
The most representative holographic duality is the AdS/CFT Correspondence (Maldacena 1998). In this duality we link gravity in a weakly curved AdS$_{5}\,\times\,S^{5}$ with a CFT in $3+1$  dimensions, which is in the conformal boundary of AdS. AdS/CFT Correspondence has strong/weak duality too, which relates SUGRA backgrounds at strong coupling with CFT at weak coupling. Thanks to this, it has been possible to construct toy models for thermal (non perturbative) QCD, as for example, the dual models of QGP using Dp/Dq branes as gravitational background.

The idea behind the AdS/CFT Correspondence is the geometrical connection between the isomeries of AdS and the conformal group. To be more precise,  since AdS is a maximally symmetric space, its isomeries are holomorphic to the Poincare Group. This implies that, at inner level, AdS and any CFT are \emph{essentially} the same thing. The statement of the correspondence is 

\begin{equation}\label{KSW}
Z_{\text{String}}\left[\phi,\mathcal{M}\right]=Z_{CFT}\left[\phi_0,\mathcal{O};\partial\,\mathcal{M},\eta\right],
\end{equation}

where $\mathcal{M}$ is the manifold where gravity lies, $\phi$ is a bulk field with $\phi_0$  as the value at the conformal boundary  $\partial \mathcal{M}$. The conformal boundary carries a metric in a fixed conformal class $[\eta]$. The conjecture stablishes that $\phi_0$ acts as a Schwinger source for any CFT operator $\mathcal{O}$ living on $\partial\,\mathcal{M}$. This is the essence of the correspondence. 

Some remarks. The conjecture in principle can be made with any background $\mathcal{M}$ that satisfies string equations of motion and has a the pair $\left(\partial\,\mathcal{M}\, , \eta \right)$. Since the solution is not unique, i.e., the charts over $M$ are not trivial, the  correlation functions are dependent from the choice of coordinates.  As a conclusion, it is possible to obtain different holographies according to the choice of chart. For example, in the Maldacena's original proposal, the Anti de Sitter space is covered partially with a Poincare chart $AdS_5$ that picks up one of the two folds of the hyperbolic space, fixing a conformal boundary at the origin of the radial coordinate of $AdS_5$. This conformal boundary has a topology of $\mathbb{R}^{1,3}$. Choices of different charts on Anti de Sitter space lead to boundaries as $\mathbb{R}^4$, $S^1\times S^{3}$, $S^1\times \mathbb{R}^3$ or $S^1 \times \mathbb{H}^3$ \cite{braga}. All of these topologies are diffeomorphical between each other. This has a deeper implication in the foundations of the correspondence, because different charts could lead to different dualities. 

The utility of the correspondence comes in the calculation procedure, encoded in the holographic dictionary, which is the relation between the bulk and the boundary physics. Since the Anti de Sitter radius and the string lenght are free parameters, it is possible to take a low energy limit in (\ref{KSW}) in order to reduce the string generating function to a supergravity one, 

\begin{equation}\label{KSW2}
W_{CFT}\left[\phi_0,\mathcal{O};\partial \, \mathcal{M}, \eta\right]=-\text{ln}\,Z_{CFT}\left[\phi_0,\mathcal{O};\partial\,\mathcal{M},\eta\right]= \sum_i Z_{\text{SUGRA}}\left[\phi,\mathcal{M}_i\right]+O\left(\frac{1}{N}\right)+O\left(\frac{1}{\sqrt{\lambda}}\right),
\end{equation}
where the sum in the supergravity action appears to take into account the chart dependence. The supergravity description  is valid only for the large $N$ and large 'tHooft coupling $\lambda$. Note that the supergravity action can carry divergences due to infinite volume or IR behaviour. These divergences must be renormalized \cite{skend1} and could lead to anomalies.

The dictionary is obtained following the saddle point approximation and the functional standard techniques from the supergravity on-shell action:
\begin{equation}\label{dict}
\langle \mathcal{O}\left(x_1\right)\,\mathcal{O}\left(x_2\right)...\mathcal{O}\left(x_n\right)\rangle_{CFT}=\left.\frac{\delta^n S_{\text{SUGRA}}^{\text{on-shell}}\left[\phi_0,...\right] }{\delta \phi_0\left(x_1\right) ...\phi_0\left(x_n\right) }\right|_{\text{Sources}=0}.
\end{equation}
Expresion (\ref{dict}) tells how to connect fields in both sides. For example, the dilaton is related with the string coupling. For each possible supergravity action a dictionary can be constructed. This is the path followed, for example, in AdS/QCD models \cite{kiritsis2}.
\section{Geometrical Approximation to the Correspondence}
\subsection{Formal Aspects}
Geometrically speaking, the correspondence is build up using the complex geometry language. Consider a open $n+1$-dimensional manifold $\left(M,g\right)$. This manifold $M$ will be the configuration space for the possible physical states on the bulk. Along with this manifold, we define a closed $n+1$-dimensional manifold $(\tilde{M}, \tilde{g})$ with no empty $n$-dimensional boundary $\partial \tilde{M}$, such that $M\subset\tilde{M}$. A complete Riemmann metric $g$ on $M$ is called  \emph{conformally compact}\footnote{Conformally compact is equivalent to Penrose compact.} if a function $f\in\,\Omega_0(\tilde{M})$ on $\tilde{M}$ exist such that 
\begin{equation}\label{conf}
\tilde{g}=f^2\,g,
\end{equation}
with $f^{-1}\left(0\right)=\partial \tilde{M}$ and $df$ is not zero on $\partial \tilde{M}$. Such a function is called a \emph{defining} function \cite{anders}. The metric $\tilde{g}$ is called \emph{compactification} of the metric $g$. The compactification defines an induced metric  $\eta =\tilde{g}|_{\partial \tilde{M}}$ on $\partial \tilde{M}$.

There are many defining functions, and hence many conformal compactifications of a given metric $g$, then the choice of $\eta$ is not unique. This problem can be avoided using the conformal class $[\eta]$ (called conformal infinity) of  $\eta$ on $\partial \tilde{M}$ defined  by conformal transformations of $\eta$. Recall that $[\eta]$
is uniquely determined by the pair $(M,g)$. Physically, the choice of $[\eta]$ implies that the causal structure of spacetime is conserved under conformal transformations. The pair $(\partial \tilde{M}, \eta)$ with $\eta \in [\eta]$,  defines the \emph{conformal boundary}, where the CFT operators are constructed. 

Since the symmetries of $M$ and $\partial\tilde{M}$ must be the same\footnote{Both manifolds must have the same causal structure.}, the moduli space of $\partial\tilde{M}$, $\mathcal{M}_{\partial\tilde{M}}$ is defined by $\mathcal{M}_{\partial \tilde{M}}=\text{Teich}(M)/\text{MCG}(M)$, since both manifolds must have the same conformal stucture because they are diffeomorphic. Following the discussions above, the entire moduli space of $M$, $\mathcal{M}$,  is restricted by the choice of a conformal class $[\eta]$, thus not all the metrics $g$ on $M$ will contribute to the partition function on the bulk. The restricted moduli space of $M$, $\mathcal{M}_{(\partial\tilde{M}, [\eta])}$, is defined as the set of all the conformally compact metrics $g$ on $M$ \cite{sanch1}. 

Under these ideas, the AdS/CFT Correspondence can be summarized saying that given any bulk data $(M,g)$, it is possible to  construct (or obtain) a boundary $(\partial\tilde{M},[\eta])$ by means of the conformal compactification scheme (\ref{conf}), i.e, 
\begin{equation}\label{conjecture}
\overbrace{Z\left(\partial\tilde{M},[\eta]\right)}^{\text{Boundary}}=\overbrace{\sum_{g\in\mathcal{M}_{(\partial\tilde{M}, [\eta])}}Z\left(g,M\right).}^{\text{Bulk}}
\end{equation}

Following physical arguments from AdS/CFT Correspondence, $M$ must be 10-dimensional.Thus, in order to have a conformal boundary as $\mathbb{R}^{(1,3)}$, $M$ has to be decomposed into $M=\mathbb{H}^5\times X^5$, with $X^5$ some compact space, such that in the compactification limit $M\backsim \mathbb{R}^{2,4}\subset \mathbb{H}^5$, as in the AdS/CFT Correspondence, in which $M$ is factorized as $AdS_5\times S^5$. All the metrics $g$ that satisfies these conditions are the so called \emph{asymptotically hyperbolic Einstein metrics}. 

\subsection{Geodesic Compactifications}
Any compactification (\ref{conf}) with a defining function given by $f_g=\text{Dist}_g\left(x,M\right)$
is called \emph{geodesic} \cite{anders,sanch1}. These compactifications are useful for computational purposes, and because given a conformal infinity $[\eta]$ of $(M,g)$ exists a unique geodesic defining function $f_{[\eta]}$ that has $\eta\in [\eta]$ as a  boundary metric. 

Following the Gauss lemma, the compactificacion $\tilde{g}$ can be expanded into
\begin{equation}\label{gl}
\tilde{g}=dr^2+g_{f}, 
\end{equation}
where $g_f$ is a family of metrics on $\partial \tilde{M}$. The Fefferman--Graham expansion \cite{feffer} of $g$ is a  truncated Taylor-type expansion of the family of metrics $g_f$, that depends on the dimensionality  $n$ of $M$. The exact form of the series depends on whether $n$ is even or odd. In a general case, the series can be written as

\begin{equation}\label{fg_exp}
g_f=g_{\left(0\right)}+r\,g_{\left(1\right)}+r^2\,g_{\left(2\right)}+\ldots +r^n\,g_{\left(n\right)}+ \text{terms depending of even or odd }n, 
\end{equation}
  
where $\eta:=g_{\left(0\right)}$ and  the coefficients $g_{\left(k\right)}$ with $1<k<n$ are locally fixed by the curvature of $\eta$ and its covariant derivatives. The extra terms depending on the even--odd character of $n$ are calculated from the Einstein equations for $\eta$.

The $g_{\left(n\right)}$ term is a little more complex. For even dimensions, $g_{\left(n\right)}$
 is transverse traceless, but is determined by global properties of $M$. In odd dimensions, $g_{\left(n\right)}$ is not traceless but is still determined by global  aspects of $M$. The $g_{\left(n\right)}$ factor corresponds to the stress--energy tensor of the CFT living in $\partial \tilde{M}$. 
 
Mathematically, these expansions can be obtained by compactifying the Einstein equations and taking Lie derivatives of $\tilde{g}$ with $f_g=0$:
\begin{equation}
g_{\left(k\right)}=\frac{1}{k!}\,\mathcal{L}_{\tilde{\nabla}f_g}^{\left(k\right)}\tilde{g}.
\end{equation}
If the metric is Hoelder, all the expansions hold up to order $m+\alpha$, with $\alpha$ the Hoelder exponent. 

As a conclusion, knowing $g_{\left(0\right)}$ and $g_{\left(n\right)}$ allow to construct the bulk metric field $g$ from the expansion (\ref{fg_exp}). The real problem here is to know the convergence of the series  and how its inclusion may introduce anomalies \cite{witten2}. 

\subsection{General Decomposition of $M$}
Until now, all of the approach to the conjecture was classical, i.e, real manifolds only. A quantum approach (thinking on strings instead of supergravity) requires a more general factorization $M=X\times Y$, where $X\in \mathbb{H}^5$   and $Y$ is a 5-dimensional Calabi--Yau manifold.  
The Calabi-Yau manifold can be justified on the grounds that classical mechanics requires a simplectic structure while quantum mechanics requires complex structure to implement unitarity. The main problem with these structures lies on the construction of Calabi-Yau metrics. This problem can be partially avoided by considering $Y$ as a 5-dimensional Sasaki-Einstein manifold \cite{sparks}.  

\section{Geometrical Issues}
As it was said above, the central idea for the construction of the conjecture is the existence of a conformal infinty $[\eta]$ that fixes a conformal boundary $\partial\tilde{M}$. This process is highly depending on the convergence of the Fefferman-Graham expansion (\ref{fg_exp}), which could introduce undesirable anomalies due to the holographic renormalization. But this is not the only problem. 

In a realistic approximation, Gravity/Gauge duality suggests that any quantum field theory must have a string dual. The large $N$ and large $\lambda$ limits restrict the possible dual models to the AdS/CFT Correspondence, that is no realistic. Leaving aside the limits, to obtain a non-conformal holography would imply the naive idea of taking a different background from Type IIB supergravity.

Advances in this scenario were given by Skenderis and Taylor with their precision holography \cite{taylor}. The idea is to categorize all the possible $X$ manifolds in the decomposition $M=X\times Y$ into spaces that are \emph{asymptotically AdS} and those which are not. Asymptotically AdS spaces are  related to the usual 10-dimensional  $AdS_{n+2}\times S^{8-n}$ through a Weyl transformation. This transformation redefines the coupling constant of the QFT on the conformal boundary including an energy scale with no trivial running. As a conclusion from \cite{taylor}, only on asymptotically AdS spaces it is possible to do non conformal holography. This implies that holographic extension only can be made on AdS-like spaces. 

Another issue arises when the index theory comes into play. Following \cite{anders,sparks}, the conjecture is build up in conformal boundaries, where the index of any pseudodifferential operator is well defined. When closed and compact manifolds  are considered, the index theorem fails. This problem leads to the consideration of the definition and the role of the boundary in AdS/CFT Correspondence \cite{sanch1}.     

\section{Conclusions}
AdS/CFT Correspondence is strongly related to the concept of conformal boundary. The construction of this boundary is dependent on the chosen charts, thus the holographic dictionary (\ref{dict})  is not univocally. The usual chart used to do holography is the Poincare chart. Non-conformal extensions are made relaxing the conformal symmetry of $AdS_5\times S^5$.

The choice of a Calabi-Yau (or a Sasaki--Einstein) manifold as the compact space in the factorization $AdS_5\times Y^5$, besides the relaxation of the large $N$ and large $\lambda$ limits could lead to string/QFT duality. 

\end{document}